\documentclass[epj]{webofc}
\usepackage[varg]{txfonts}   % Web of Conferences font
\wocname{EPJ Web of Conferences}
\woctitle{INPC 2013}
\begin{document}
\title{Exploring the QCD phase diagram through relativistic heavy ion collisions}
%
% subtitle is optional

\author{Bedangadas Mohanty\inst{1}\fnsep\thanks{\email{bedanga@niser.ac.in}} 
        % etc.
}

\institute{School of Physical Sciences, National Institute of Science Education and Research, Bhubaneswar 751005, India
          }

\abstract{%
We present a review of the studies related to establishing the QCD phase diagram through high energy nucleus-nucleus
collisions. We particularly focus on the experimental results related to the formation of a quark-gluon phase,
crossover transition and search for a critical point in the QCD phase diagram.
}%
\maketitle
\section{Introduction}
\label{intro}

Physical systems can be made to undergo phase transitions by varying parameters such as the
temperature ($T$) or a chemical potential ($\mu$) of the system. Systems whose underlying interactions
are strong interactions, are not different. In the theory of strong interactions, Quantum
Chromodynamics (QCD), there are distinct conserved quantities. For a grand canonical ensemble of strongly 
interacting particles, the conserved baryon, electric charge and strangeness numbers are associated 
with the corresponding chemical potentials $\mu_{B}$, $\mu_{Q}$, and $\mu_{S}$, respectively.  
So for a system with strong interactions one can lay out the phase diagram with axes being 
$T$, $\mu_{B}$, $\mu_{Q}$, and $\mu_{S}$. Experimentally such a system of strong interactions 
can be created by colliding two nuclei at high energy. However, in such a system one can only vary 
to an appreciable extent $T$ and $\mu_{B}$ (values of $\mu_{Q}$, and $\mu_{S}$ are small~\cite{Abelev:2008ab}). 
This can be done by varying the center of mass energies ($\sqrt{s_{\rm {NN}}}$) of the 
collision of the two heavy nuclei~\cite{Cleymans:2005xv,BraunMunzinger:2007zz}. Hence through 
relativistic heavy-ion collisions we can explore a two dimensional phase diagram,  
$T$ versus $\mu_{B}$, of strong interactions~\cite{Rajagopal:2000wf}.

Such a phase diagram has several distinct phase structures. Some of which are: 
(a) high temperature and/or density phase of deconfined quarks and gluons (QGP),
(b) low temperature and/or density phase of hadrons, (c) nature of quark-hadron transition
is crossover for the small $\mu_{B}$ part of the phase diagram and 
first order for the rest (large $\mu_{B}$) of the phase diagram, and 
(d) end point of the first order phase transition line (called the critical point (CP)). 

In this paper, we will present the experimental results that support the formation
of a deconfined phase of quarks and gluons in high energy nuclear collisions,
experimental and theoretical evidences towards establishing the quark-hadron
transition as a crossover and progress towards the search for the critical point
on the phase diagram. 

\section{Quark-Gluon Phase}
\label{sec-1}

The experimental programs at the Relativistic Heavy Ion Collider (RHIC) facility~\cite{Adams:2005dq,Adcox:2004mh,Gyulassy:2004zy} and 
the Large Hadron Collider (LHC) facility~\cite{Singh:2013fha} have conclusively provided evidences towards
the formation of a deconfined state of matter in heavy-ion collisions where the relevant 
degrees of freedom are quarks and gluons. In this section we will only discuss a subset
of these results. We will also show that these signatures get gradually turned-off when 
the beam energy is dialled down to low enough values of $\sqrt{s_{\rm {NN}}}$ 
(RHIC Beam Energy Scan (BES) Program~~\cite{Abelev:2009bw,Mohanty:2011nm}), where temperature and energy 
density achieved are possibly below those predicted by QCD calculations for a 
quark-hadron transition. Typical temperature and energy density for phase transition are
of the order of 170 MeV and 1 GeV/$fm^3$, respectively~\cite{Karsch:2004ti}.

\subsection{Nuclear Modification Factor}
\label{sec-11}
\begin{figure}[h]
\centering
\includegraphics[width=10cm]{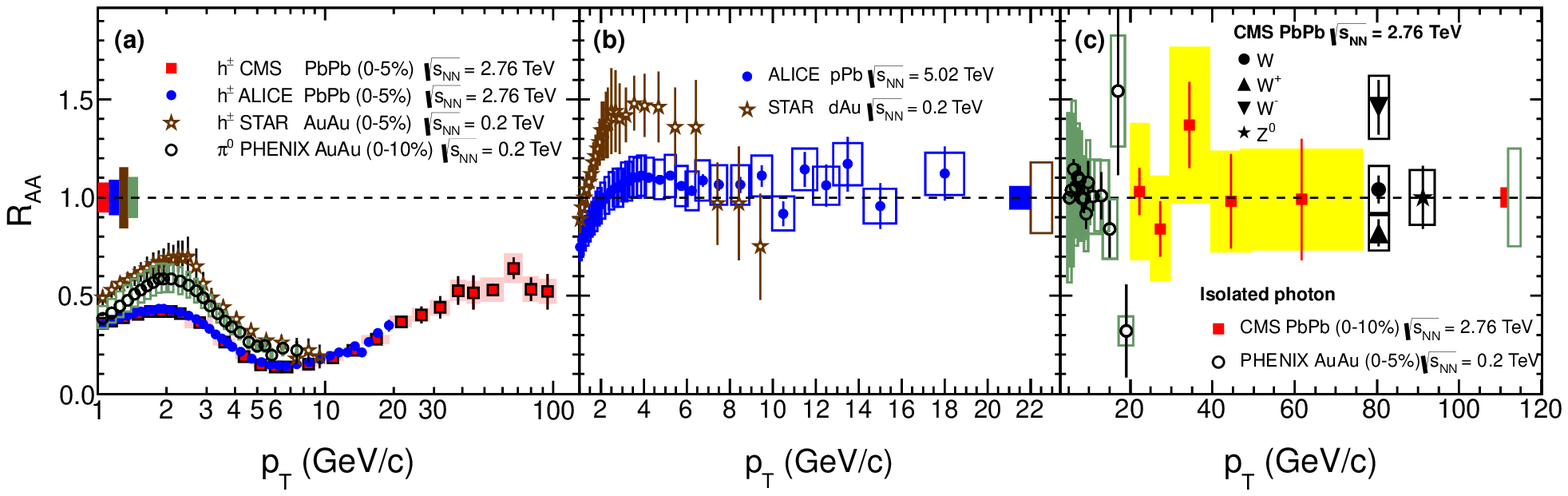}
\includegraphics[width=5cm]{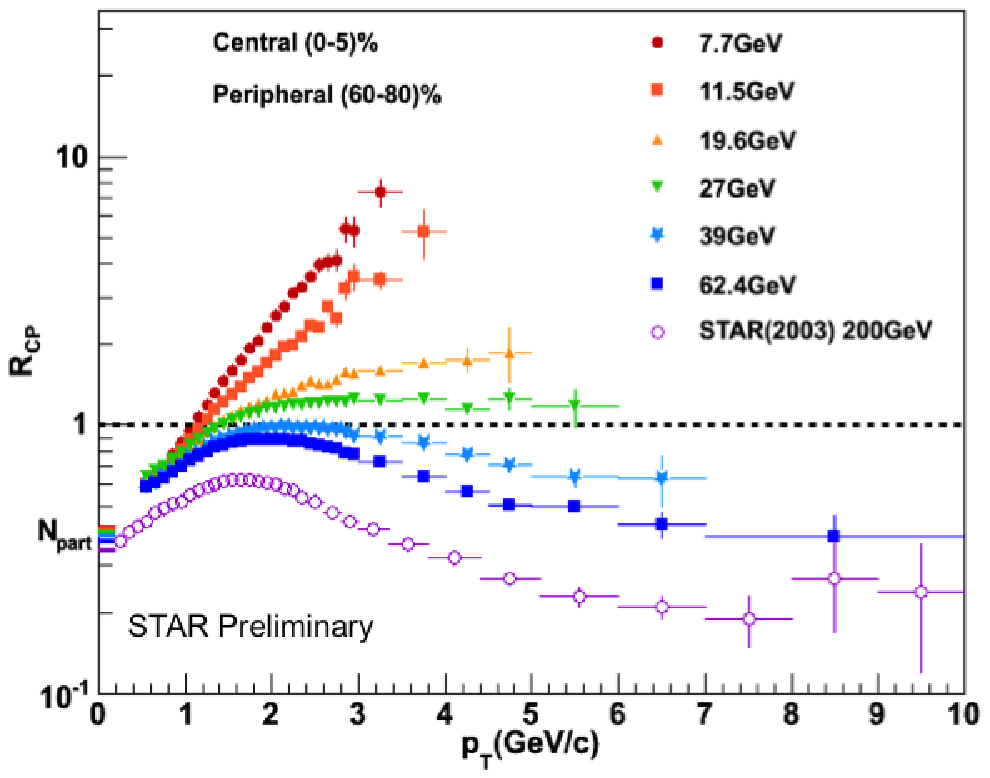}
\vspace{-0.5 cm}
\caption{
(Color online) Top panel: (a) Nuclear modification factor $R_{\rm {AA}}$ of charged hadrons measured 
by LHC experiments (ALICE~\cite{Aamodt:2010jd} and CMS~\cite{CMS:2012aa}) and 
$R_{\rm {AA}}$ of hadrons measured by RHIC experiments (charged hadrons by 
STAR~\cite{Adams:2003im} and $\pi^{0}$ by PHENIX~\cite{Adler:2006hu}).
(b) Comparison of nuclear modification factor for charged hadrons versus $p_{T}$ 
for minimum bias collisions in $d$+Au collisions at RHIC~\cite{Adams:2003im} 
and $p$+Pb collisions at LHC~\cite{ALICE:2012mj}. 
(c) The $R_{\rm {AA}}$  versus $p_{T}$ for isolated photons in central nucleus-nucleus collisions at 
RHIC~\cite{Adler:2005ig} and LHC~\cite{Chatrchyan:2012vq}. Also shown are the $R_{\rm {AA}}$ of 
$W^{\pm}$~\cite{Chatrchyan:2012nt} and $Z$ bosons~\cite{Chatrchyan:2011ua} at LHC energies.
The boxes around the data denote $p_{T}$-dependent systematic uncertainties.
The systematic uncertainties on the normalisation are shown as boxes at $R_{\rm {AA}}$ = 1.
Bottom panel:  Nuclear modification factor ($R_{\rm {CP}}$ ) versus transverse momentum for charged hadrons in RHIC BES program~\cite{Horvat:2013rga}.}
\label{raa}       % Give a unique label
\end{figure}

The nuclear modification factor ($R_{\rm {AA}}$) is defined as 
$\frac{dN_{AA}/d\eta d^2 p_{T}} {T_{AB} d\sigma_{NN}/d\eta d^2
  p_{T}}$, where $T_{AB} = N_{binary}/\sigma_{inelastic}^{pp}$  is the overlap 
integral (with $N_{binary}$ being the number of binary collisions
commonly estimated from Glauber model calculation) 
and $d\sigma_{NN}/d\eta d^2 p_{T}$
is the cross section of charged hadron production in $p$+$p$ collisions.  $R_{\rm {AA}}$ at high transverse
momentum ($p_{T}$) having a value of less than one is attributed to energy loss of partons in QGP and this
phenomenon is referred to as the jet quenching in a dense partonic matter~\cite{Wang:1991xy}. 

Figure~\ref{raa} shows the  $R_{\rm {AA}}$ of various particles produced in nucleus-nucleus collisions at RHIC and LHC.
In Fig.~\ref{raa}(a), we observe that it is $R_{\rm {AA}}$ $<$ 1 both
at RHIC and LHC. 
In Fig.~\ref{raa}(b), we observe that the nuclear modification factors for $d$+Au collisions at
$\sqrt{s_{\mathrm {NN}}}$~=~200 GeV~\cite{Adams:2003im} and  $p$+Pb collisions at  
$\sqrt{s_{\mathrm {NN}}}$ = 5.02 TeV~\cite{ALICE:2012mj} are greater
than unity for $p_{T}$ $>$ 2 GeV/$c$ both at RHIC and LHC. 
We do not expect formation of a dense partonic matter in such collisions. The nuclear modification factor value in $p(d)$+A
collisions being close to unity suggests that $p(d)$+A collisions can be considered 
as due to superposition of several $p$+$p$ collisions. It acts as an experimental support to the view that a hot and 
dense medium of color charges is formed in nucleus-nucleus collisions at RHIC and LHC.  In Fig.~\ref{raa}(c), 
we show the $R_{\rm {AA}}$ of particles that do not participate in strong interactions. 
These particles (photon~\cite{Adler:2005ig,Chatrchyan:2012vq}, 
$W^{\pm}$~\cite{Chatrchyan:2012nt} and $Z$~\cite{Chatrchyan:2011ua}  bosons) have a $R_{\rm {AA}}$ $\sim$ 1, 
indicating that the $R_{\rm {AA}}$ $<$ 1, observed for hadrons in nucleus-nucleus collisions, are due to the strong 
interactions in a dense medium consisting of color charges.  Going down in $\sqrt{s_{\rm {NN}}}$ gradually increases
the value of $R_{\rm {CP}}$ (ratio of yields of charged particles in
central to peripheral collisions normalized to respective $N_{binary}$
) at high $p_{\rm T}$ towards unity and eventually becomes larger than unity for 
 $p_{\rm T}$ $>$ 2 GeV/$c$ for $\sqrt{s_{\rm {NN}}}$ =  11.5 and 7.7 GeV~\cite{Horvat:2013rga}.
This indicates a turn-off of QGP signature at lower beam energies, although $p$+A collisions at these energies are
required to quantitatively confirm the findings.

\subsection{Partonic Collectivity}
\label{sec-12}

%%%%%%%%%%%%%%% Fig.2  %%%%%%%%%%%%%%%%%%%%%%%%%%%%%
\begin{figure}
\begin{center}
\includegraphics[scale=0.3]{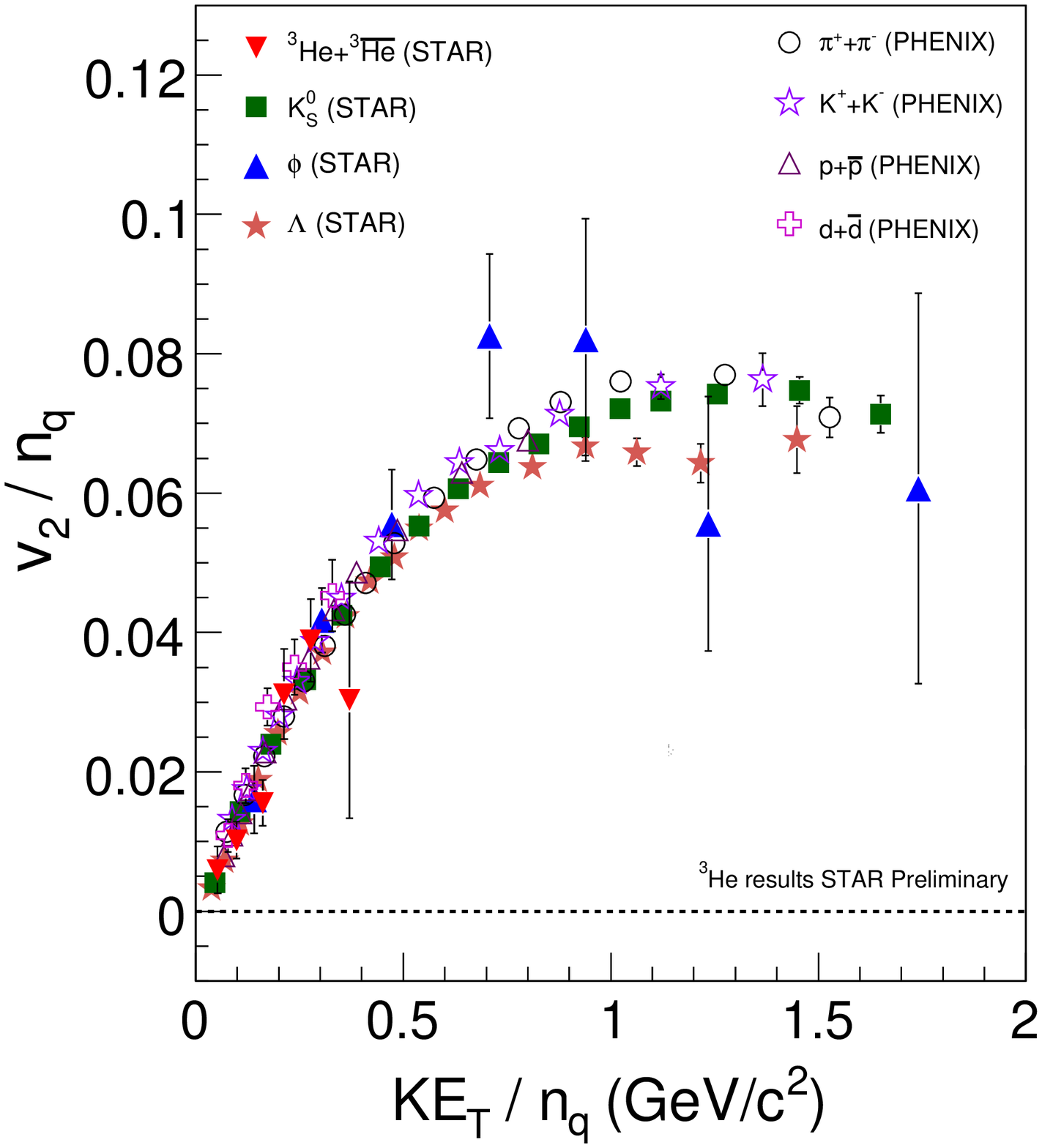}
\includegraphics[scale=0.4]{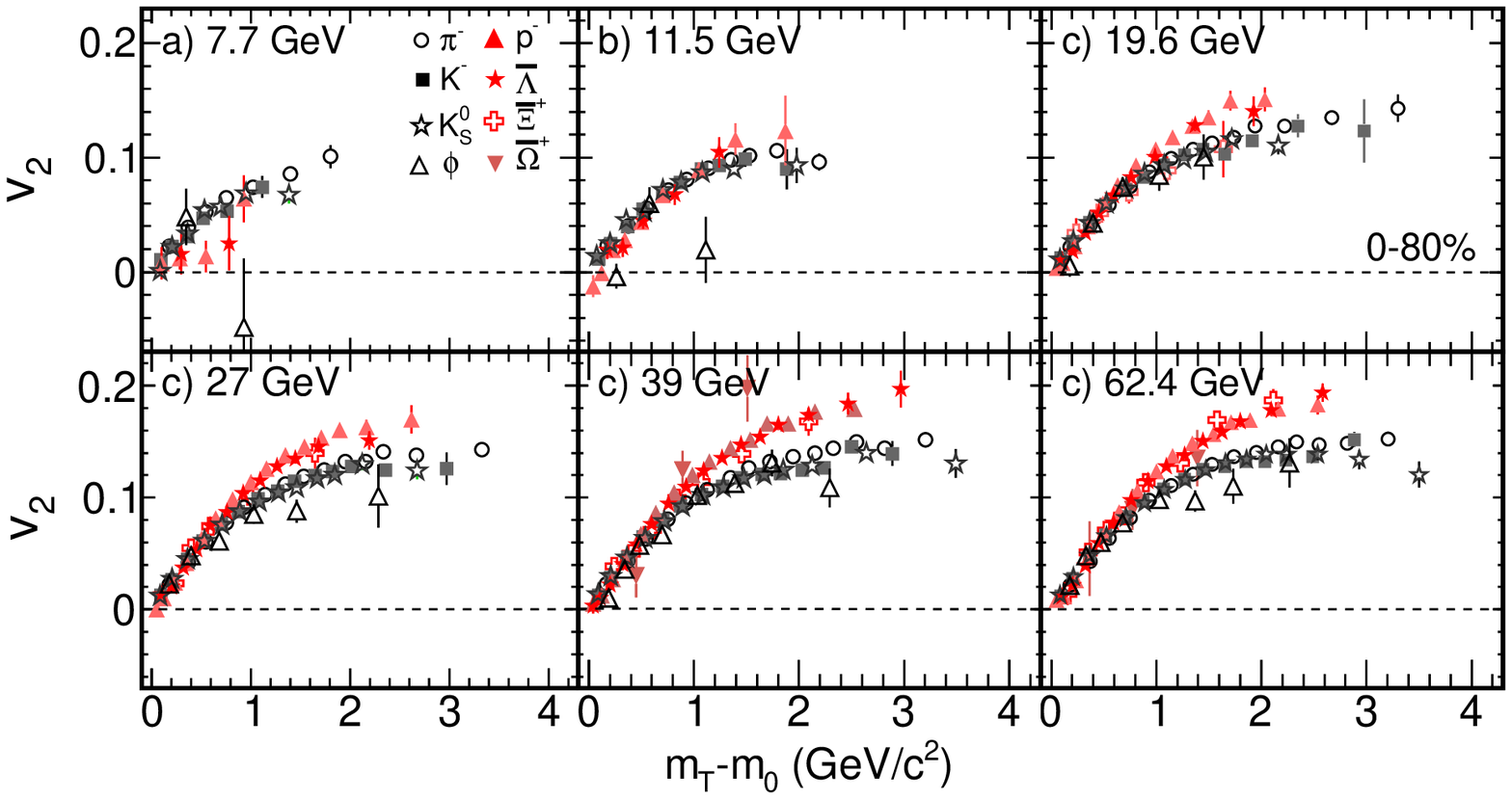}
\vspace{-0.5 cm}
\caption{(Color online) Left panel: Elliptic flow ($v_{2}$) divided by 
$n_{q}$ versus  ($m_{T}$ - $m$)/$n_{q}$ (denoted as $KE_{T}/n_{q}$) for various produced particles 
in Au+Au collisions at $\sqrt{s_{\rm {NN}}}$ = 200 GeV~\cite{Mohanty:2011ki}. Right panel: 
$v_{2}$ versus $m_{T}$ - $m_{0}$ for various produced particles in RHIC BES program~\cite{Adamczyk:2013gw}.}
\label{v2}
\end{center}
\end{figure}
%%%%%%%%%% End of Fig. 2  %%%%%%%%%%%%%%%%%%%%%%%%%%%%%

The elliptic flow ($v_{2}$) is calculated as $\langle cos~2(\phi - \Psi_{\rm 2}) \rangle$, 
where $\phi$ denotes the azimuthal angle of the produced particles 
and $\Psi_{\rm 2}$ denotes the orientation of the second order event plane (plane subtended by the impact parameter and
beam axis). A new observation
was made at RHIC in the measurement of $v_{2}$ for various hadrons. For  $p_{\rm T}$ $>$ 2 GeV/$c$,
there as a clear splitting of $v_{2}$ observed on the basis of baryons and mesons. All the baryons
were found to have similar $v_{2}$ and in turn all the mesons were to found to have similar $v_{2}$ which
was lower compared to the baryons. Dividing the  $v_{2}$ by the number of constituent quarks ($n_{q}$ 
= 2 for mesons and 3 for baryons) and plotting it versus ($m_{T}$ - $m_{0}$)/$n_{q}$ (where $m_{T}$
is the transverse mass and $m_{0}$ is the rest mass of the hadron) showed a remarkable scaling for most of the
measured hadrons and light nuclei (shown in Fig.~\ref{v2} left panel)~\cite{Mohanty:2011ki}. All these particles have different interaction 
cross sections, freeze-out at different times and have large variations in their masses, yet
they show a similar $v_{2}$/$n_{q}$ when plotted versus ($m_{T}$ - $m_{0}$)/$n_{q}$. This indicates
that a substantial amount of the measured $v_{2}$ is developed in the partonic phase and the
contribution from hadronic phase is small~\cite{Abelev:2007rw,Afanasiev:2007tv}. 
Hence these measurements support the idea of formation
of a medium with quarks and gluons. Fig.~\ref{v2} (right panel) shows that the baryon-meson splitting 
(for $m_{T}$ - $m_{0}$ $>$ 1 GeV/$c^2$ and source of this partonic collectivity) reduces gradually as the beam energy is decreased and
vanishes for $\sqrt{s_{\rm {NN}}}$ =  11.5 and 7.7 GeV~\cite{Adamczyk:2013gw}. This indicates a turn-off of this 
QGP signature at lower beam energies, where the collision process perhaps does not create a matter
of sufficiently high temperature and density.

\subsection{Dynamical Charge Correlations}
\label{sec-13}

%%%%%%%%%%%%%%% Fig.3  %%%%%%%%%%%%%%%%%%%%%%%%%%%%%
\begin{figure}
\begin{center}
\includegraphics[scale=0.4]{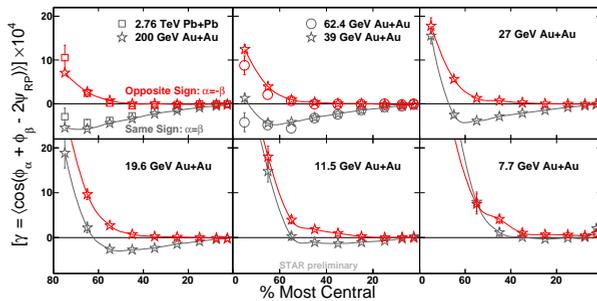}
\vspace{-0.5 cm}
\caption{(Color online) Charge hadron azimuthal correlations with respect to reaction plane angle 
as a function of centrality for nucleus-nucleus collisions at midrapidity\cite{Wang:2012qs}.} 
\label{lpv}
\end{center}
\end{figure}
%%%%%%%%%% End of Fig. 3  %%%%%%%%%%%%%%%%%%%%%%%%%%%%%

Figure~\ref{lpv} shows the results on charged hadron azimuthal correlations based on
3-particle correlation technique\cite{Abelev:2009ac}. The results are from nucleus-nucleus collisions 
at $\sqrt{s_{\mathrm {NN}}}$ = 7.7,11.5, 19.6, 27, 39, 64.2, 200 GeV~\cite{Wang:2012qs,Abelev:2009ac}  and 2.76 TeV~\cite{Abelev:2012pa} at midrapidity
between same--charge and opposite--charge hadrons with respect to the reaction plane angle ($\Psi$).
The observable, $\langle \cos(\phi_\alpha +\phi_\beta -2\psi) \rangle$ 
represents the difference between azimuthal correlations (between two particles $\alpha$ and $\beta$) projected
onto the direction of the angular momentum vector and correlations projected
onto the collision event plane. 
The difference between the same charge and opposite charge correlations 
at the higher energies seems to be consistent with the predictions for
existence of metastable domains in QCD vacuum leading to local {\it parity} violation.
This phenomena needs deconfinement and chiral phase transitions which are expected to be
achieved in heavy-ion collisions (also referred to as the Chiral Magnetic Effect (CME))~\cite{Fukushima:2008xe,Kharzeev:2004ey}. 

We also observe that the difference in correlations between same and opposite charges seems to decrease 
as beam energy decreases and almost vanishes at 7.7 GeV. If the differences can be attributed
to QCD transitions, absence of it may indicate absence of such transitions
at the lower energies. The observable presented is parity-even, making it also susceptible 
to physical processes not related to CME.

\section{Order of Phase Transition}
\label{sec-2}

\begin{figure}
%\begin{center}
\includegraphics[scale=0.35]{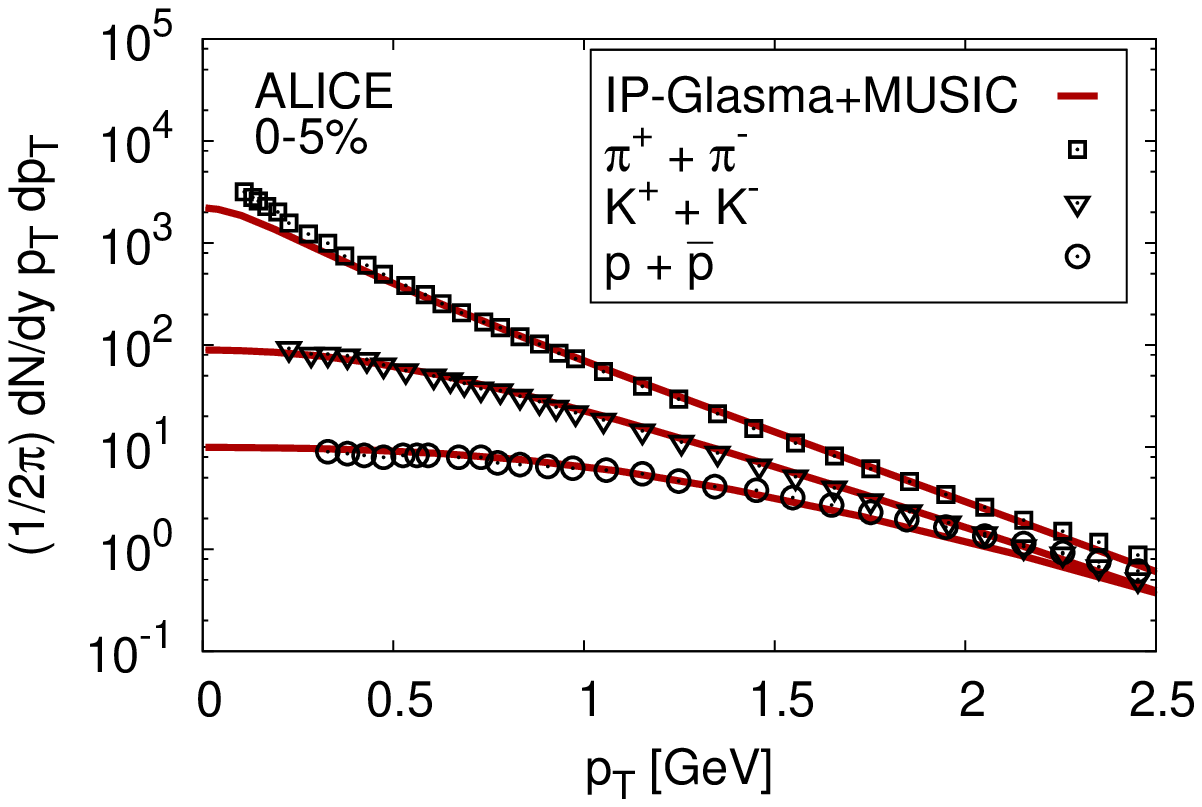}
\includegraphics[scale=0.35]{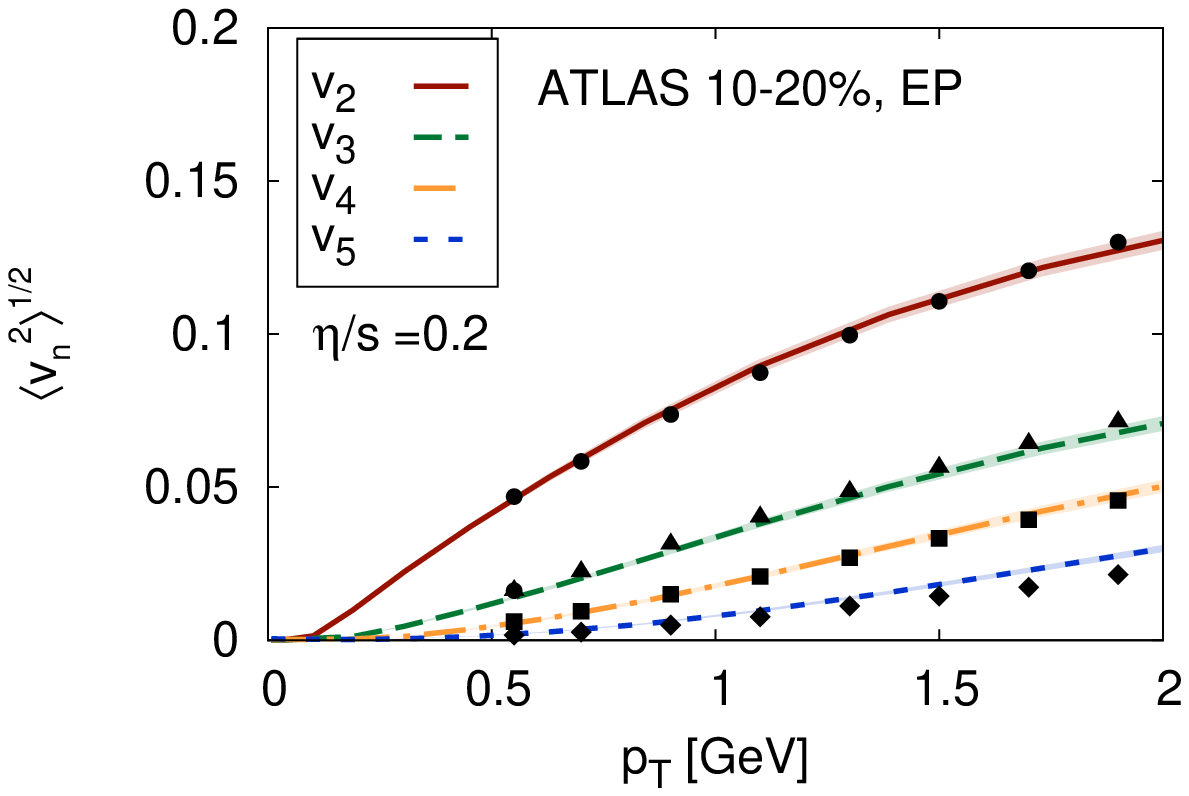}
\includegraphics[scale=0.25]{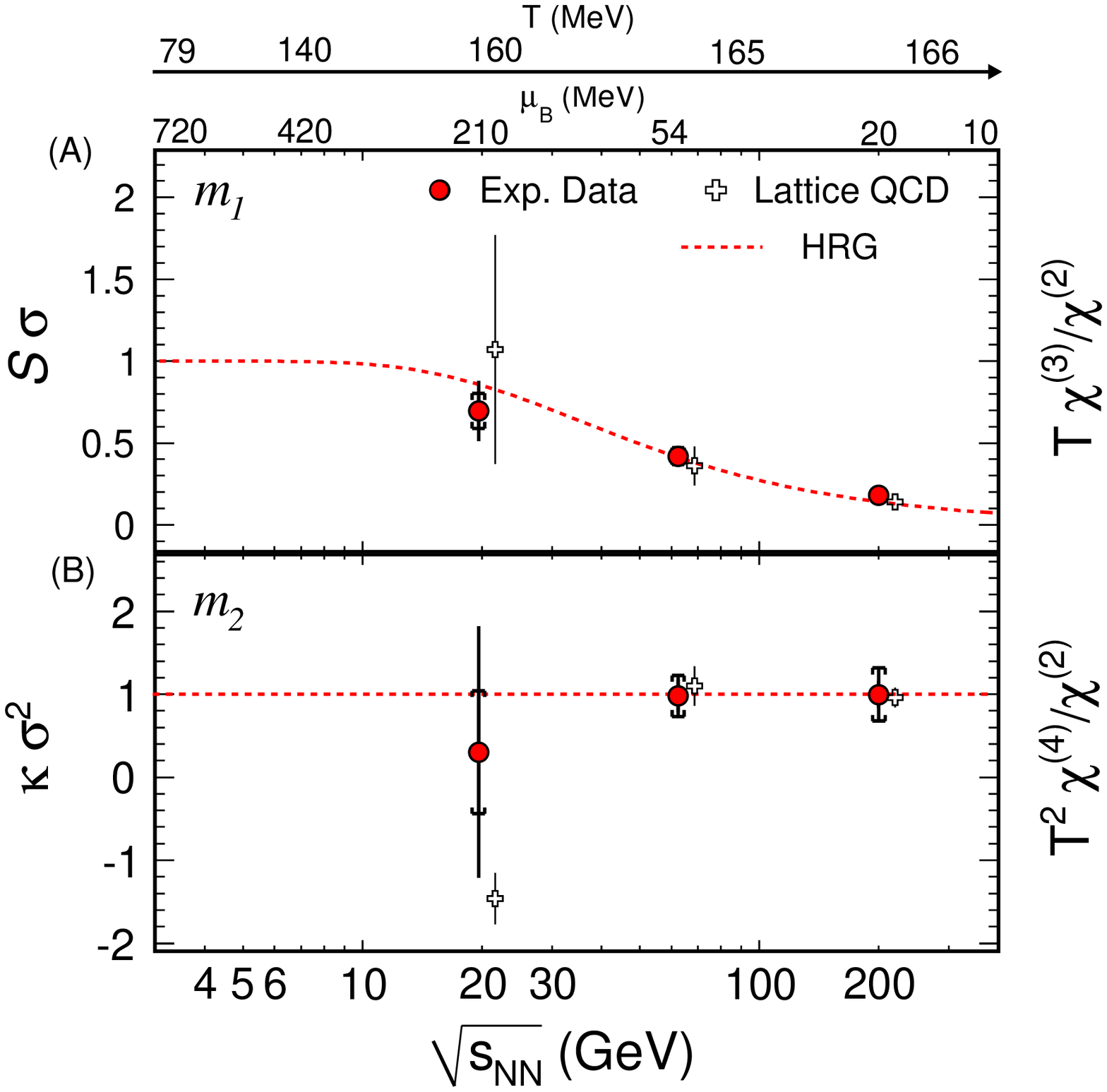}
\vspace{-0.5 cm}
\caption{\label{fig:co}
(Color online) Left panel: Hydrodynamical calculation of identified hadron transverse momentum
spectra compared to experimental data from the ALICE collaboration, Middle panel: Root-mean-square
anisotropic flow co-efficients as a function of transverse momentum compared to experimental
data by the ATLAS collaboration~\cite{Gale:2012rq}, and 
Right panel: Comparison of lattice QCD and experimental data at RHIC. 
Experimentally measured ratios of cumulants of net-proton distributions, $S\sigma$ and $\kappa\sigma^{2}$,
where $\sigma$, $S$ and $\kappa$ are the second, third and fourth moments,  are shown as
a function of $\sqrt{s_{\mathrm {NN}}}$ for impact parameter values of less than 3 fm for
Au+Au collisions at RHIC \cite{Gupta:2011wh}. Also plotted on the top scale are the chemical freeze-out
values of $\mu_B$ and $T$ corresponding to $\sqrt{s_{\mathrm {NN}}}$ as obtained from a
hadron resonance gas model, which assumes the system to be in chemical
and thermal equilibrium at freeze-out \cite{Cleymans:2005xv,BraunMunzinger:2007zz}. }
%\end{center}                                                                                                                        
\end{figure}

First principle QCD calculations on lattice at high temperature and $\mu_{B}$ = 0 MeV 
have established the quark-hadron transition to be a crossover~\cite{Aoki:2006we}.
The lattice chiral susceptibility $\chi(N_s,N_t)$ = $\partial^2$/$(\partial$$m_{ud}^2)$($T/V$)$\cdot\log Z$, 
as a function of temperature was found to be independent of volume of the system. $m_{ud}$ is the mass of the
light u,d quarks, $N_s$ is the spatial extension, $N_\tau$ the euclidean time extension,  and $V$ the system volume.

Using this crossover equation of state for the quark-hadron transition
in a hydrodynamic based model, the experimental data on invariant yields of charged hadrons and
various order azimuthal anisotropy as a function of transverse momentum at LHC are
nicely explained (shown in Fig~\ref{fig:co})~\cite{Gale:2012rq}. Lending indirect support to the transition
being a crossover at small $\mu_{B}$. In addition, the ratio of baryon susceptibilities ($\chi^{n}$) of 
various orders ($n$) computed in lattice QCD calculations with a crossover at $\mu_{B}$ = 0 MeV, when compared 
to experimental measurements of similar quantities using correlations between measured protons and 
anti-protons (reflected by the product of moments of net-proton distribution) also shows a remarkable
agreement within errors.

The crossover temperature is obtained by looking for the point of sharpest change in temperature dependence of the
chiral susceptibility ($\chi_{\bar{\psi}\psi}$), the strange quark number susceptibility ($\chi_s$)
and the renormalized Polyakov-loop ($L$) in the lattice calculations.
Various lattice QCD estimates of chiral crossover temperature using $\chi_{\bar{\psi}\psi}$
have converged to a value of around 154 MeV~\cite{Aoki:2009zzc,Bazavov:2011nk}.
However, the observables ($\chi_s$ and $L$) that provide important insights into deconfining aspect of the crossover
show a slightly higher crossover temperature of around 170 MeV. But with a width of around 15 MeV in 
temperature estimates, it is difficult to make a concerte statement on the difference between various measures
of the crossover temperatures. Moreover there are unresolved discussions on the establishment 
of Polyakov loop expectation and strange quark number susceptibilities to critical behaviour 
in the light quark mass regime~\cite{Bazavov:2011nk}.

\section{Critical Point}
\label{sec-3}

\begin{figure}
\begin{center}
\includegraphics[scale=0.5]{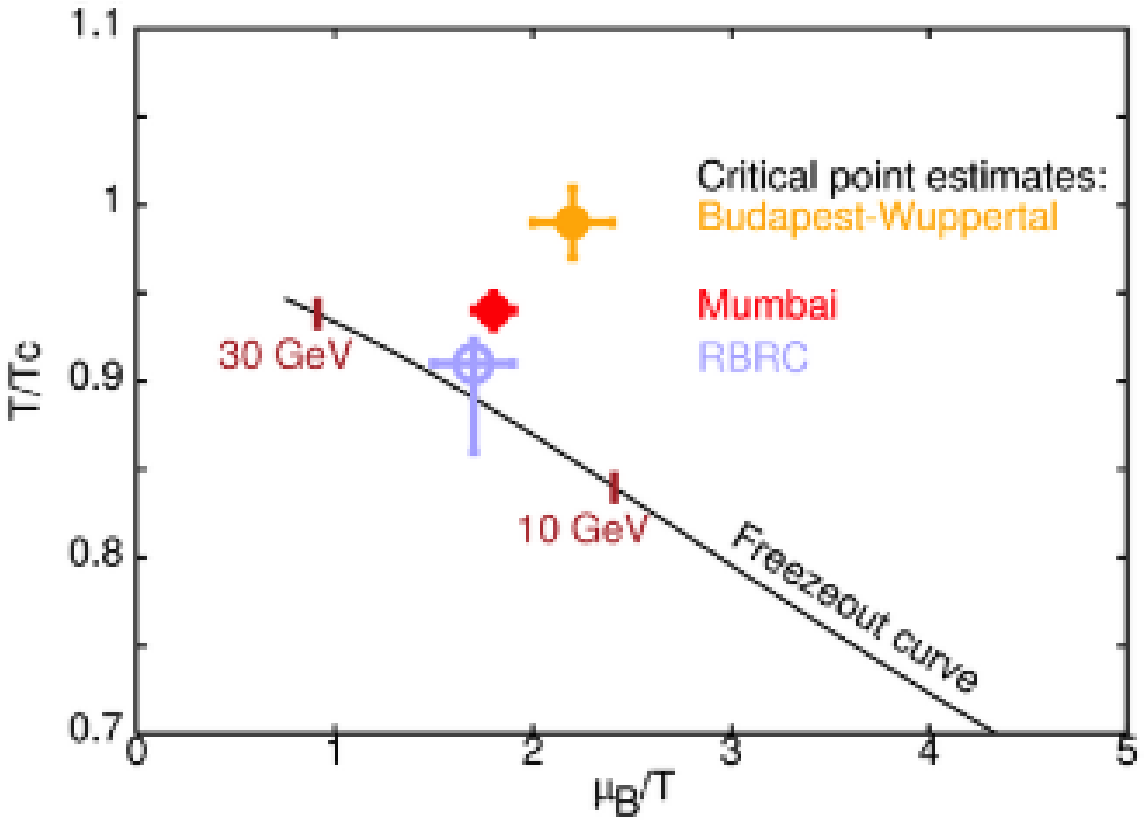}
\includegraphics[scale=0.5]{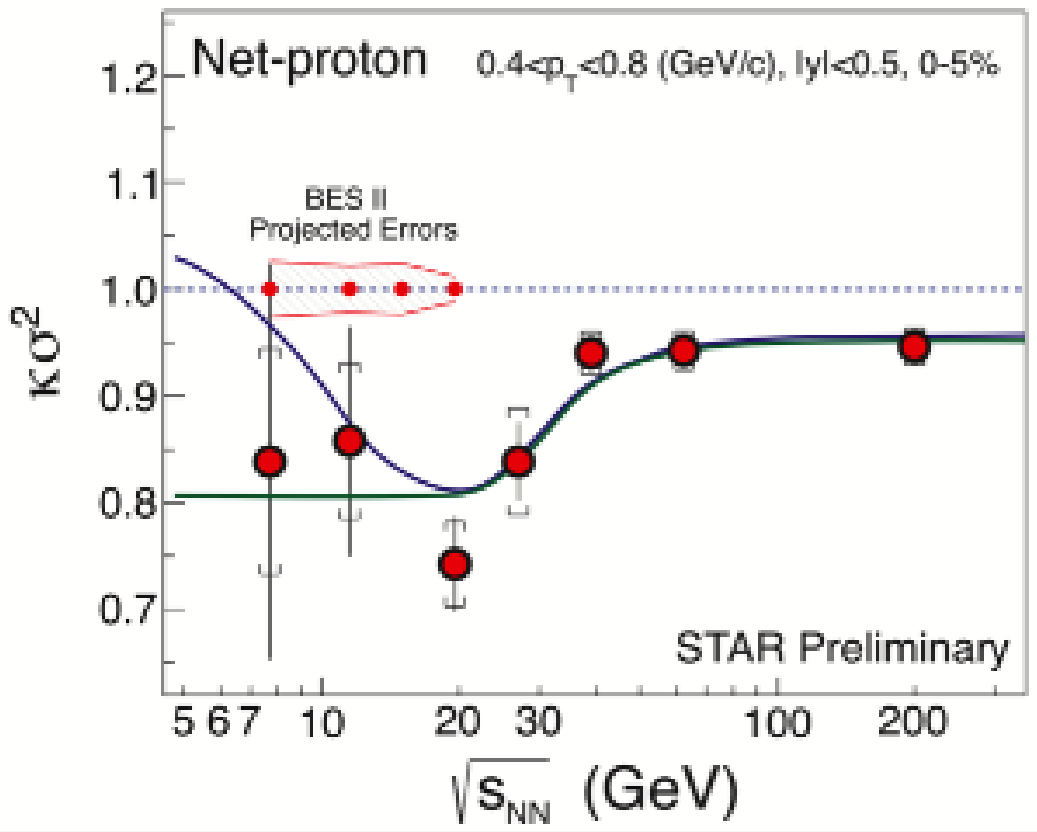}
\vspace{-0.5 cm}
\caption{\label{fig:cp}
(Color online) Left panel: Estimates of position of critical point from various lattice QCD calculations~\cite{Gupta:2011zzd,Gupta:2012zze}.
Right panel: $\kappa \sigma^2$ for net-proton distributions as a function of $\sqrt{s_{\mathrm {NN}}}$ in RHIC BES
program~\cite{Luo:2013saa}. Also shown are the projected statistical error in the second phase of BES program.}
\end{center}
  \end{figure}

Several QCD based models predict the existence of an end point or critical point (CP) at high $\mu_{B}$
for the first order phase transition in the QCD phase diagram. However the exact
location depends on the model assumptions used~\cite{Stephanov:2004wx}. Given the
ambiguity in predictions of CP in models, studies on lattice were expected to provide reliable estimates~\cite{Gavai:2004sd}.
However lattice calculations at finite $\mu_{B}$ are difficult due to {\it sign problem}.
There are several ways suggested to overcome this issue. (i) Reweighting the partition function
in the vicinity of transition temperature and $\mu$ = 0~\cite{Fodor:2004nz}, (ii) Taylor expansion of thermodynamic
observables in $\mu$/T about $\mu$ = 0~\cite{Gavai:2003mf}  and (iii) Choosing the chemical potential to be imaginary
will make the fermionic determinant positive~\cite{Philipsen:2009yg}.  The first two methodologies yield an
existence of CP, whereas the third procedure gives a CP only when the first coefficient in the
Taylor expansion of generic quark mass on the chiral critical surface ($m_{c}$) as a function of
$\mu$/T ($\frac{m_c(\mu)}{m_c(0)} = 1+\sum_{k=1} c_{k} \left(\frac{\mu}{\pi T_c}\right)^{2k}$)
is positive. The lattice calculations which yield a CP on phase diagram are shown 
in Fig.~\ref{fig:cp}~\cite{Gupta:2011zzd,Gupta:2012zze}.
However these calculations still have to overcome some of the lattice
artefacts like lattice spacing,
physical quark masses, volume effect and continuum limit extrapolation.

%----------------------------------------------------------------------------------------------------                               
  %----------------------------------------------------------------------------------------------------                                                                

In the experimental side, the characteristic signature of CP is large fluctuations in event-by-event
conserved quantities like net-charge, net-baryon number and net-strangeness. The
variance of these distributions ($\langle (\delta N)^{2} \rangle$) are proportional to square
of the correlation length ($\xi$). It has been shown that higher moments ($\langle (\delta N)^{3} \rangle$
$\sim$ $\xi^{4.5}$ and $\langle (\delta N)^{4} \rangle$ $\sim$ $\xi^{7}$)
have stronger dependence on $\xi$ compared to variance and hence have 
higher sensitivity~\cite{Stephanov:2008qz,Stephanov:2011pb,Asakawa:2009aj}.
In addition the moments are related to  susceptibilities~\cite{Cheng:2008zh}.
Motivated by all these, experiments are studying the variable $\kappa$$\sigma^2$ of net-proton
distributions (a proxy for net-baryon, see caption of Fig.~\ref{fig:cp}),  to search for the CP.
$\kappa$$\sigma^2$ will be constant as per the Central limit theorem and hadron resonance gas model~\cite{Garg:2013ata}.
They have monotonic dependence with $\sqrt{s_{\mathrm {NN}}}$ for non-CP scenarios~\cite{Luo:2013bmi}. 
However as it is related to the ratio of baryon number susceptibilities in QCD models:
$\kappa$$\sigma^2$ = $\frac{\chi^{(4)}_{\mathrm B}}{\chi^{(2)}_{\mathrm B}/T^2}$~\cite{Gavai:2010zn},
close to  CP it is expected to show a non-monotonic dependence on  $\sqrt{s_{\mathrm {NN}}}$.
Preliminary experimental results on  $\kappa$$\sigma^2$ value for net-proton distributions
measured in RHIC BES program are shown in the right panel of Fig.~\ref{fig:cp}~\cite{Luo:2013saa}. Interesting trends
are observed indicating that the CP if exists in the phase diagram, have to be below
$\sqrt{s_{\mathrm {NN}}}$ = 39 GeV~\cite{Aggarwal:2010wy}.

\section{Summary}
\label{sec-4}

Relativistic heavy-ion collision experiments have seen distinct signatures which suggest 
that the relevant degrees of freedom at top RHIC and LHC energies in the initial stages 
of the collisions are quark and gluons and the system quickly approaches thermalization. 
The underlying mechanism for the fast thermalization is currently under study. 
Lowering the beam energies to 11.5 GeV and below leads to a smooth turning-off of 
the QGP signatures, indicating that hadronic interactions dominate.
These observations in turn further support the formation of partonic matter at higher 
energy collisions. Three such QGP signatures related to phenomena of jet-quenching,
partonic collectivity and charge correlations (or chiral magnetic effect) are 
discussed in this paper.

The QCD calculations on lattice tell us that the above observed quark-hadron transition at LHC and
top RHIC energies (small values of $\mu_{B}$ $\sim$  0 MeV) is a crossover. The theoretical evidence
lies in the chiral susceptibility versus temperature not changing with change in volume. The 
chiral crossover temperature is found to be around 154 MeV. Other observables of quark-hadron 
crossover give a slightly higher values of the crossover temperature with large uncertainties.
Using the lattice QCD based crossover equation of state in hydrodynamic models, one could explain the 
various measurements at RHIC and LHC. Two such measurements related to the transverse momentum 
distribution and azimuthal distribution of the produced particles in nucleus-nucleus collisions 
are presented in this paper. These lend indirect support from experimental measurements that 
the observed quark-hadron transition at LHC and top RHIC energies is a crossover. 

Most QCD calculations on lattice continue to indicate the possible existence of critical point 
for $\mu_{B}$ $>$ 160 MeV, this possibility has not been ruled out from the data at RHIC. 
The exact location is not yet known unambiguously. The experimental measurements though 
encouraging are inconclusive. High event statistics measurements in the second phase of 
RHIC beam energy scan program should be able to provide a more quantitative and hence conclusive
picture from the experimental side. Computing intensive lattice QCD calculations 
removing the remaining artefacts pertaining to lattice spacing, quark masses, system volume and
choice of action will clear the picture from the theoretical side. 

The progress towards establishing the phase diagram of strong interactions, one of the four
basic interactions that occur in nature through relativistic heavy-ion collisions is 
significant. The phase structures of quark-gluon and hadronic phase has been distinctly 
identified, the transition is a crossover, with crossover temperatures varying between
150 - 175 MeV (depending on the observable used), and both experiment and theory have 
ruled out the existence of the critical point for $\mu_{B}$ $<$ 160 MeV in the QCD 
phase diagram.

Acknowledgement: Financial support for the work is obtained from the DST Swarna Jayanti Fellowship.

% BibTeX or Biber users please use (the style is already called in the class, ensure that the "woc.bst" style is in your local directory)
% \bibliography{name or your bibliography database}

\end{document}